\documentclass[lettersize,journal]{IEEEtran}
\usepackage{amsmath,amsfonts}
\usepackage{algorithmic}
\usepackage{algorithm}
\usepackage{array}
\usepackage[caption=false,font=normalsize,labelfont=sf,textfont=sf]{subfig}
\usepackage{textcomp}
\usepackage{stfloats}
\usepackage{url}
\usepackage{verbatim}
\usepackage{graphicx}
\usepackage{bbm}
\usepackage{mathtools}
\usepackage{subfig}
\usepackage{overpic}
\usepackage{booktabs}
\usepackage{multirow}
\usepackage{svg}
\usepackage{enumitem}
\usepackage{siunitx}
\usepackage{makecell}
\usepackage[hidelinks]{hyperref}
\usepackage{xcolor}
\usepackage{xpatch}

\makeatletter
\def\changeBibColor#1{%
	\in@{#1}{9392366, misra2016cross, 9348261, lin2023libmtl, 9887407, 9632584, 10227799}
}
\xpatchcmd\@bibitem
{\item}
{\changeBibColor{#1}\item}
{}{}

\makeatother
\usepackage{xcolor}
\begin{document}

\title{CSPMNet: Pareto-Efficient Automatic Modulation Classification With Learnable Complex Subband Phase Motion}

\author{Ruixiang Zhang, Zinan Zhou, Yezhuo Zhang, Guangyu Li and Xuanpeng Li
\thanks{This work has been submitted to the IEEE for possible publication. Copyright may be transferred without notice, after which this version may no longer be accessible. \textit{Corresponding author: Xuanpeng Li}}
\thanks{Ruixiang Zhang, Zinan Zhou, Yezhuo Zhang and Xuanpeng Li are with the School of Instrument Science and Engineering, Southeast University, Nanjing, 210096, Jiangsu, China (e-mail: \protect\url{zhang_ruixiang@seu.edu.cn}; \protect\url{zhouzinan919@seu.edu.cn}; \protect\url{zhang_yezhuo@seu.edu.cn}; \protect\url{li_xuanpeng@seu.edu.cn}).}
\thanks{Guangyu Li is with the School of Computer Science and Engineering, University of Science and Technology, Nanjing, 210094, Jiangsu, China (email: guangyu.li2017@njust.edu.cn).}
\thanks{Digital Object Identifier XX.XXXX/LWC.2024.XXXXXXX}
}

\markboth{Journal of \LaTeX\ Class Files,~Vol.~14, No.~8, August~2021}%
{Shell \MakeLowercase{\textit{et al.}}: A Sample Article Using IEEEtran.cls for IEEE Journals}

\IEEEpubid{0000--0000/00\$00.00~\copyright~2021 IEEE}

\maketitle
\begin{abstract}
Automatic modulation classification (AMC) is an essential technique for noncooperative spectrum monitoring and intelligent wireless receivers.
However, practical AMC models must identify modulation formats from short and noisy I/Q observations while maintaining low computational and storage overhead.
Existing deep-learning approaches often improve recognition accuracy by expanding generic neural backbones, which increases deployment cost and weakens their suitability for resource-constrained receivers.
To bridge the gap between recognition performance and model efficiency, this letter proposes a Complex Subband Phase-Motion Network, designated as CSPMNet, for lightweight AMC from raw I/Q samples.
Specifically, learnable complex subband filters are introduced to adaptively extract frequency-selective baseband responses while preserving the algebraic coupling between in-phase and quadrature components.
Then, an amplitude-preserving phase-motion module captures multi-lag temporal rotation dynamics within each subband, and a lightweight temporal classifier performs efficient sequence aggregation.
Rigorous experimental evaluations on public RadioML benchmark datasets demonstrate that CSPMNet achieves highly competitive recognition accuracy while requiring substantially lower model complexity than many existing AMC models.
Codes are available on \href{https://github.com/klay7w/CSPMNet}{GitHub}.
\end{abstract}


\section{Introduction}
\IEEEPARstart{A}{utomatic} modulation classification (AMC) enables a noncooperative receiver to infer the modulation format of an observed I/Q sequence without transmitter-side metadata \cite{peng2021survey}.
This function is required in cognitive radio, spectrum monitoring, and interference analysis, where unknown signals must be identified before demodulation or protocol-level processing.
In practical sensing receivers, the observation is often short and noisy, and the classifier must remain small enough for repeated spectrum sensing. 
Although recent deep AMC models \cite{li2026iqcm} enhance recognition by expanding network capacity to learn from raw I/Q sequences or statistical features, sustaining satisfactory classification robustness within a stringent computational budget remains a critical bottleneck.

This difficulty primarily stems from the representation burden placed on compact classifiers \cite{zhang2021efficient, zhang2023toward}.
To reliably distinguish diverse modulation formats, an AMC model must simultaneously capture distinct spectral characteristics, amplitude variations, and phase dynamics from short baseband $I/Q$ sequences \cite{chen2024multi, zhang2023amc}.
While scaling up network capacity helps learn these coupled physical factors, it often leads to high-complexity or even multi-million-parameter backbones that are ill-suited for edge sensing \cite{zhang2021efficient}.
Conversely, merely shrinking the network size to lower computational costs often expels crucial modulation-discriminative details \cite{zhang2023toward}.
Therefore, to break this Pareto dilemma, an efficient AMC model should shift the representation burden to a physics-guided front-end, which explicitly injects complex-valued algebraic constraints and stabilizes noise-sensitive physical domains before decision-making.

\begin{figure}[h]
    \centering
    \includegraphics[width=\linewidth]{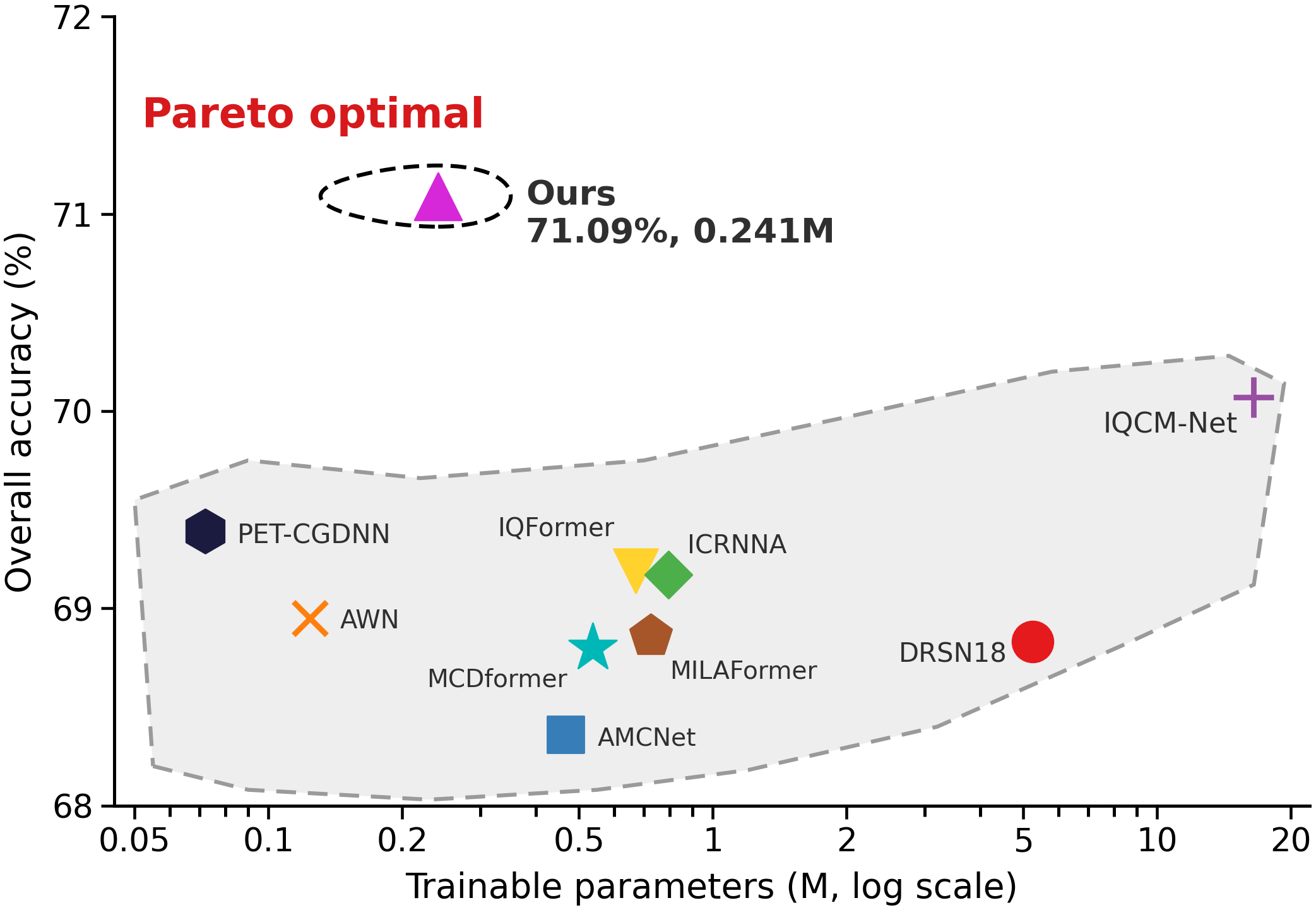}
    \caption{Accuracy-parameter tradeoff on RadioML2022.01A.}
    \label{fig:pareto}
\end{figure}

\IEEEpubidadjcol

To implement such a physics-guided front-end, this letter introduces CSPMNet, a learnable Complex Subband Phase-Motion Network designed for Pareto-efficient AMC.
As quantified by the accuracy-parameter frontier visualized in Fig.~\ref{fig:pareto}, CSPMNet improves the accuracy-parameter tradeoff.
Specifically, under a rigorous benchmark protocol on RadioML2022.01A \cite{sathyanarayanan2023rml22}, CSPMNet achieves an overall accuracy of 71.09\% with a mere 0.241M trainable parameters, creating a distinctly favorable Pareto operating point that outperforms both parameter-heavy and existing lightweight networks.
By incorporating the inherent physical priors into the network front-end, CSPMNet reduces the representation pressure on subsequent neural layers.
The main contributions of this letter are as follows:
\begin{itemize}[topsep=0pt, partopsep=0pt, itemsep=0pt, parsep=0pt]
\item[$\bullet$] We introduce learnable complex subband into AMC from raw I/Q samples. This design replaces fixed transform preprocessing with task-adaptive complex filtering, preserving I/Q coupling and exposing frequency-selective signal components before lightweight recognition.
\item[$\bullet$] We delve into amplitude-aware phase-motion modeling on the learned subbands. By constructing multi-lag complex phase-motion products together with magnitude reliability, the proposed representation avoids relying on phase-only angular cues and retains temporal rotation information that magnitude-only summaries may discard.
\item[$\bullet$] Our CSPMNet demonstrates strong Pareto-efficient performance. Compared with reproduced benchmark models, it delivers higher recognition accuracy with substantially fewer trainable parameters, and maintains robust performance across low-, mid-, and high-SNR regimes.
\end{itemize}

\section{Problem Description}
For a modulated signal with modulation type $m$, the received continuous-time waveform is modeled as
\begin{equation}
r(t)=\psi_m(s(t))*p(t)+n(t),
\end{equation}
where $s(t)$ denotes the transmitted symbol, $\psi_m(\cdot)$ is the modulation mapping associated with class $m$, $p(t)$ is the channel impulse response, $n(t)$ is additive white Gaussian noise, and $*$ denotes convolution.
After down-conversion and sampling, each received sample is represented by an I/Q sequence
\begin{equation}
x =
\begin{bmatrix}
I[0], I[1], \ldots, I[T-1]\\
Q[0], Q[1], \ldots, Q[T-1]
\end{bmatrix}
\in \mathbb{R}^{2 \times T},
\end{equation}
where $I[\cdot]$ and $Q[\cdot]$ are the in-phase and quadrature-phase components. 
In the benchmarks used in this letter, $T=128$.

Let $\mathcal{D}=\{(x_i,y_i)\}_{i=1}^{N}$ denotes a labeled AMC dataset, where $y_i$ belongs to the candidate modulation set $\mathcal{C}=\{c_1,c_2,\ldots,c_C\}$. 
AMC aims to learn a classifier $h_{\theta}: \mathbb{R}^{2 \times T}\rightarrow \mathbb{R}^{C}$, where $h_{\theta}(x)_c$ denotes the predicted score for modulation class $c$. The predicted label is then obtained by
\begin{equation}
\hat{y}=\arg\max_{c\in\mathcal{C}} h_{\theta}(x)_c .
\end{equation}
This letter considers lightweight AMC, where the expected classification loss should be minimized under a limited model budget:
\begin{equation}
\min_{\theta}\ \mathbb{E}_{(x,y)}[\ell(h_{\theta}(x),y)],
\quad \mathrm{s.t.}\quad P(\theta)\leq B,
\end{equation}
where $\ell(\cdot)$ is the classification loss, $P(\theta)$ denotes the number of trainable parameters, and $B$ is the parameter budget.
Under a small $B$, the classifier has limited capacity to learn rich modulation-discriminative features, motivating an explicit front-end for lightweight classification.

\section{Proposed Method}
\subsection{Overall Architecture of CSPMNet}
\begin{figure*}[t]
    \centering
    \includegraphics[width=0.8\linewidth]{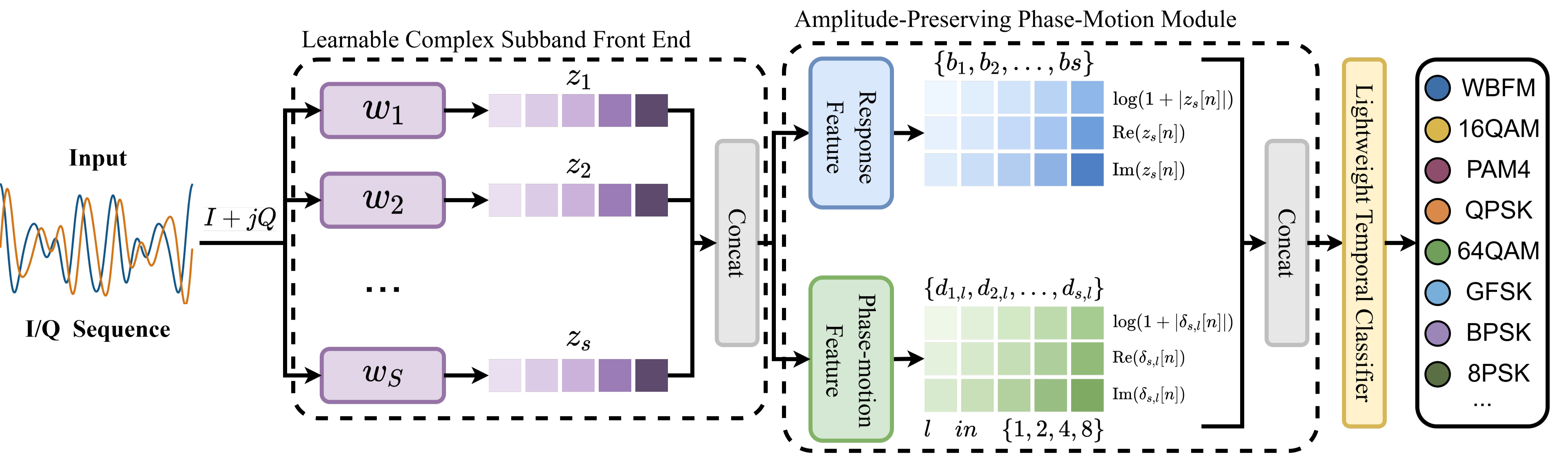}
    \caption{Architecture of CSPMNet.}
    \label{fig:architecture}
\end{figure*}

To reduce the representation burden, CSPMNet first transforms the raw I/Q sequence into a signal-informed feature map and then performs lightweight temporal aggregation.
As shown in Fig.~\ref{fig:architecture}, the framework contains three consecutive components: a learnable complex subband front end, an amplitude-preserving subband phase-motion module, and a lightweight temporal classifier.
The central design principle is to expose modulation-relevant complex dynamics before the final classifier, rather than relying on a large generic backbone to learn feature implicitly.

Given the input $x \in \mathbb{R}^{2 \times T}$, the first component generates a bank of complex subband responses by trainable complex filtering.
The second component computes base response features and multi-lag phase-motion products within each subband while preserving amplitude reliability.
The resulting sequence is finally fed into a lightweight classifier for modulation prediction.
Unlike fixed-transform preprocessing followed by a heavy classifier, these components are trained jointly for the AMC objective, which allows the front end to adapt to modulation-discriminative subbands while keeping the downstream classifier modest.

\subsection{Learnable Complex Subband Front End}
The first component aims to separate frequency-selective signal structures without discarding the algebraic relation between the in-phase and quadrature components.
Instead of treating the two input channels as unrelated real-valued features, CSPMNet applies $S$ trainable complex filters to the baseband sequence.
For subband $s$, the real and imaginary responses are
\begin{equation}
y_{s,r} = x_r * w_{s,r} - x_i * w_{s,i},
\quad
y_{s,i} = x_r * w_{s,i} + x_i * w_{s,r},
\end{equation}
where $x_r$ and $x_i$ denote the in-phase and quadrature components of the input sequence, $*$ denotes one-dimensional convolution, and $w_{s,r}$ and $w_{s,i}$ are the real and imaginary components of the $s$-th complex filter, respectively.
This operation follows the multiplication rule of complex convolution, so the real and imaginary parts are mixed as a coupled complex response rather than as two independent feature streams.
Such a constraint is useful for AMC because modulation differences are often expressed through joint amplitude-phase behavior in the I/Q plane.
The learned subband response is written as
\begin{equation}
z_s[n] = y_{s,r}[n] + j y_{s,i}[n]
\end{equation}
and serves as a complex subband envelope for subsequent phase-motion extraction.

Compared with fixed wavelet or scattering filters, the free complex filters can adapt their passbands to the target AMC objective.
Compared with unconstrained real-valued feature extraction, however, the front end still retains a signal-processing form: each filter produces a complex subband signal, and all subsequent phase-motion terms are computed from this response.
This design therefore combines task adaptivity with an interpretation close to the original baseband representation.

\subsection{Amplitude-Preserving Phase-Motion Module}
After subband filtering, the second component describes how the complex response evolves over time.
AMC requires both amplitude and phase dynamics.
Amplitude evolution is discriminative in itself, since envelope and subband-energy patterns help separate constant-envelope modulations from amplitude-varying ones and reflect constellation-level changes.
Meanwhile, phase increments and rotations characterize temporal angular motion, but their reliability depends on local signal energy under noise.
This is particularly important under low-SNR conditions.
If a phase descriptor is normalized to a unit-magnitude angular feature, a weak noise-dominated response may be assigned the same strength as a high-energy signal component.
Conversely, a magnitude-only descriptor may suppress the direction and rotation information that separates modulation families.
Therefore, CSPMNet constructs subband phase-motion features without removing amplitude reliability.

For each subband response $z_s[n]$, the response features are
\begin{equation}
b_s[n] = [\log(1 + |z_s[n]|), \operatorname{Re}(z_s[n]), \operatorname{Im}(z_s[n])].
\end{equation}
For each lag $l$ in $\{1, 2, 4, 8\}$, the complex phase-motion product is
\begin{equation}
\delta_{s,l}[n] = z_s[n] \operatorname{conj}(z_s[n-l]),
\end{equation}
from which the model extracts
\begin{equation}
d_{s,l}[n] = [\log(1 + |\delta_{s,l}[n]|),
              \operatorname{Re}(\delta_{s,l}[n]),
              \operatorname{Im}(\delta_{s,l}[n])].
\end{equation}
The product $\delta_{s,l}[n]$ measures the complex rotation between two samples separated by lag $l$.
The lag set $\{1,2,4,8\}$ covers both short-range and longer-range temporal dynamics within each subband.
By concatenating $b_s[n]$ and all $d_{s,l}[n]$ across the eight subbands, CSPMNet obtains a feature map at each time index.
The classifier therefore receives not only the direction of subband phase motion, but also the magnitude of the underlying response and of the phase-motion product.
This amplitude-preserving formulation distinguishes the proposed representation from phase-only angular descriptors and prevents unreliable low-energy phase estimates from being treated as equally confident evidence.

\subsection{Lightweight Temporal Classifier}
The final component aggregates the subband phase-motion sequence into a modulation prediction.
Because the preceding modules already expose frequency-selective and amplitude-aware phase dynamics, the classifier is designed to remain compact rather than to act as a large feature extractor.
The feature map is first normalized by \texttt{BatchNorm1d} and mixed by a one-dimensional convolution.
A one-layer bidirectional GRU then models temporal dependencies in the phase-motion sequence, and scaled additive attention pools the sequence by assigning larger weights to more discriminative time positions.
The representation is finally passed to a small MLP classifier.

This lightweight head matches the overall goal of Pareto-efficient AMC.
The convolution performs local channel mixing, the bidirectional GRU captures short temporal context, and the attention layer replaces uniform averaging with data-dependent temporal pooling.
Together with the small subband front end, the full CSPMNet contains 240,972 trainable parameters.
Thus, the proposed model concentrates capacity on modulation-oriented representation construction while remaining much smaller than large backbones.

\section{Experimental Results and Analysis}
\subsection{Implementation Details}
The main evaluation is conducted on RadioML2022.01A \cite{sathyanarayanan2023rml22}, and the secondary evaluation is conducted on RadioML2016.10B \cite{o2016radio}.
The detailed dataset parameters are listed in Table~\ref{tab:dataset-details}.
RadioML2022.01A contains 11 modulation classes and 21 SNR points, while RadioML2016.10B contains 10 modulation classes and 20 SNR points.

\begin{table}[t]
\caption{Detailed Parameters of Datasets}
\label{tab:dataset-details}
\centering
\small
\setlength{\tabcolsep}{0pt}
\renewcommand{\arraystretch}{1.10}
\begin{tabular*}{\columnwidth}{@{\extracolsep{\fill}}lcc@{}}
\hline
\textbf{Parameter} & \makecell{\textbf{RadioML2022.01A}} & \makecell{\textbf{RadioML2016.10B}} \\
\hline
Modulation number & 11 & 10 \\
Signal length & 128 & 128 \\
SNR range & $-20$:2:20 dB & $-20$:2:18 dB \\
Samples per mod. & 42,000 & 20,000 \\
\hline
\end{tabular*}
\end{table}

All models are implemented in Python 3.10 with PyTorch 2.4.1 and CUDA 12.4, and trained on NVIDIA GeForce RTX 3080 GPUs using the Adam optimizer.
Unless otherwise stated, all results use a fixed per-modulation per-SNR split with an effective train/validation/test ratio of 60/20/20, seed 42, 50 training epochs, batch size 512, and learning rate $1\times10^{-3}$. 
Baselines are reproduced under the same split protocol, including DRSN18\cite{zhao2019deep}, ICRNNA\cite{el2025improved}, IQFormer\cite{shao2024iqformer}, AMCNet\cite{zhang2023amc}, AWN\cite{zhang2023toward}, IQCM-Net\cite{li2026iqcm}, MCDformer\cite{chen2024multi}, MILAFormer\cite{zhao2026milaformer} and PET-CGDNN\cite{zhang2021efficient}.
Trainable parameter counts are computed from the local model implementations.

\subsection{Experiment Results}
\subsubsection{Accuracy-Complexity Comparison}
Table~\ref{tab:performance-complexity} compares CSPMNet with baselines in terms of trainable parameters, FLOPs, overall accuracy (OA) and SNR-segment accuracy.

\begin{table*}[t]
\caption{Performance and Complexity Comparison on Benchmark Datasets}
\label{tab:performance-complexity}
\centering
\small
\setlength{\tabcolsep}{2.4pt}
\renewcommand{\arraystretch}{1.08}
\begin{tabular}{@{}l|cc|cccc@{}}
\hline
\makebox[0.18\textwidth][l]{\textbf{Model}} &
\multicolumn{2}{c|}{\makebox[0.24\textwidth][c]{\textbf{Complexity}}} &
\multicolumn{4}{c}{\makebox[0.50\textwidth][c]{\textbf{Performance Metrics (\%)}}} \\
\cline{2-3}\cline{4-7}
& \makebox[0.12\textwidth][c]{\smash{\raisebox{-0.72\normalbaselineskip}{Params (M)}}}
& \makebox[0.12\textwidth][c]{\smash{\raisebox{-0.72\normalbaselineskip}{FLOPs (M)}}}
& \makebox[0.105\textwidth][c]{\smash{\raisebox{-0.72\normalbaselineskip}{OA}}}
& \makebox[0.13\textwidth][c]{\strut Low SNR}
& \makebox[0.13\textwidth][c]{\strut Mid SNR}
& \makebox[0.13\textwidth][c]{\strut High SNR} \\
\cline{5-7}
& & & & \makebox[0.13\textwidth][c]{$\leq -10$ dB}
& \makebox[0.13\textwidth][c]{$-8$--0 dB}
& \makebox[0.13\textwidth][c]{$\geq 2$ dB} \\
\hline
\multicolumn{7}{c}{\textbf{RadioML2022.01A}} \\
\hline
PET-CGDNN & \textbf{0.072} & 16.24 & 69.39 & 26.30 & 69.40 & 95.24 \\
AWN & \underline{0.124} & \underline{11.97} & 68.95 & 26.24 & 68.00 & 95.04 \\
IQFormer & 0.670 & 89.07 & 69.19 & 25.58 & 68.58 & 95.67 \\
ICRNNA & 0.795 & \textbf{10.78} & 69.17 & 25.95 & 68.93 & 95.23 \\
MCDformer & 0.537 & 198.80 & 68.80 & 26.46 & 68.82 & 94.20 \\
MILAFormer & 0.726 & 80.50 & 68.86 & 26.58 & 68.38 & 94.47 \\
AMCNet & 0.466 & 92.64 & 68.36 & 25.11 & 68.19 & 94.40 \\
DRSN18 & 5.250 & 176.96 & 68.83 & 26.34 & 68.41 & 94.54 \\
IQCM-Net & 16.482 & 848.39 & \underline{70.07} & \underline{26.93} & \underline{70.53} & \underline{95.73} \\
\textbf{CSPMNet (Ours)} & 0.241 & 49.98 & \textbf{71.09} & \textbf{28.25} & \textbf{71.94} & \textbf{96.37} \\
\hline
\multicolumn{7}{c}{\textbf{RadioML2016.10B}} \\
\hline
PET-CGDNN & \textbf{0.072} & 16.24 & 60.84 & 16.42 & 62.66 & 89.44 \\
AWN & \underline{0.124} & \underline{11.97} & 61.97 & 15.59 & 66.19 & 90.53 \\
IQFormer & 0.670 & 89.07 & 62.48 & \textbf{17.20} & 63.97 & 91.83 \\
ICRNNA & 0.795 & \textbf{10.78} & 60.24 & 14.39 & 63.26 & 89.13 \\
MCDformer & 0.537 & 198.80 & 62.37 & 15.08 & \underline{67.89} & 90.83 \\
MILAFormer & 0.726 & 80.50 & 63.15 & \underline{16.91} & 66.65 & 92.03 \\
AMCNet & 0.466 & 92.64 & 57.88 & 14.96 & 58.93 & 85.91 \\
DRSN18 & 5.250 & 176.96 & 61.18 & 14.82 & 64.61 & 90.17 \\
IQCM-Net & 16.482 & 848.39 & \underline{63.19} & 16.67 & 65.94 & \textbf{92.67} \\
\textbf{CSPMNet (Ours)} & 0.241 & 49.98 & \textbf{63.26} & 15.18 & \textbf{68.35} & \underline{92.49} \\
\hline
\end{tabular}

\vspace{0.3em}
{\scriptsize FLOPs are estimated using PyTorch FlopCounterMode under input shape $1\times2\times128$; all models are profiled with the same script and environment.}
\end{table*}

On RadioML2022.01A, CSPMNet achieves the best OA of 71.09\% and also ranks first in the low-, mid-, and high-SNR averages. 
Compared with IQCM-Net, the strongest reproduced baseline in OA, CSPMNet improves OA by 1.02 percentage points while reducing the parameter count from 16.482M to 0.241M and FLOPs from 848.39M to 49.98M. 
This corresponds to about 68.4$\times$ fewer parameters and 17.0$\times$ fewer FLOPs. 
Compared with the lightweight AWN baseline, CSPMNet improves OA by 2.14 percentage points and increases the low-, mid-, and high-SNR averages by 2.01, 3.94, and 1.33 percentage points, respectively, while still keeping the model below 0.25M parameters and 50M FLOPs.

CSPMNet is not the model with the absolute minimum computational cost. 
ICRNNA, AWN and PET-CGDNN have lower FLOPs or fewer parameters. 
However, their OA values on RadioML2022.01A remain below 69.40\%, and their low-, mid-, and high-SNR averages are all lower than those of CSPMNet. 
Thus, the proposed model does not simply trade accuracy for compactness; rather, it uses a moderate computational budget to move beyond existing lightweight baselines while avoiding the cost of a large cross-modal backbone.

On RadioML2016.10B, CSPMNet achieves comparable overall accuracy to the strongest reproduced baselines while using substantially fewer parameters.
Compared with IQCM-Net, it gives a small OA gain of about 0.08 percentage points while retaining the same 68.4$\times$ parameter reduction and 17.0$\times$ FLOP reduction.
Its high-SNR average remains close to IQCM-Net, whereas its low-SNR average is not the best.
Overall, the RadioML2016.10B results support the efficiency and competitiveness of CSPMNet, rather than implying uniform dominance across all SNR regimes.

\subsubsection{SNR-Wise Analysis}

\begin{figure*}[t]
    \centering
    \begin{minipage}[t]{0.49\textwidth}
        \centering
        \includegraphics[width=0.8\linewidth]{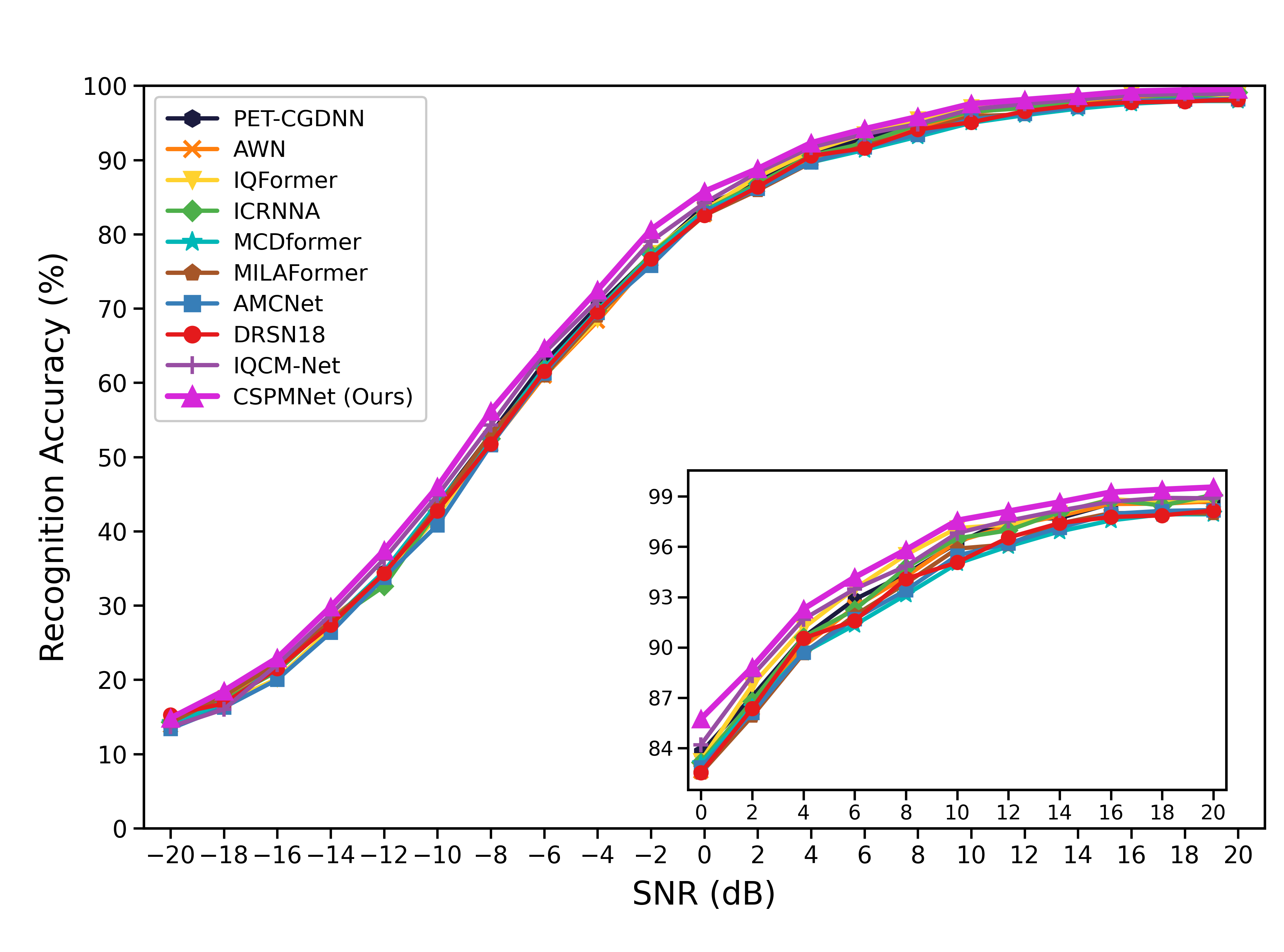}
        \centerline{(a) RadioML2022.01A}
    \end{minipage}
    \hfill
    \begin{minipage}[t]{0.49\textwidth}
        \centering
        \includegraphics[width=0.8\linewidth]{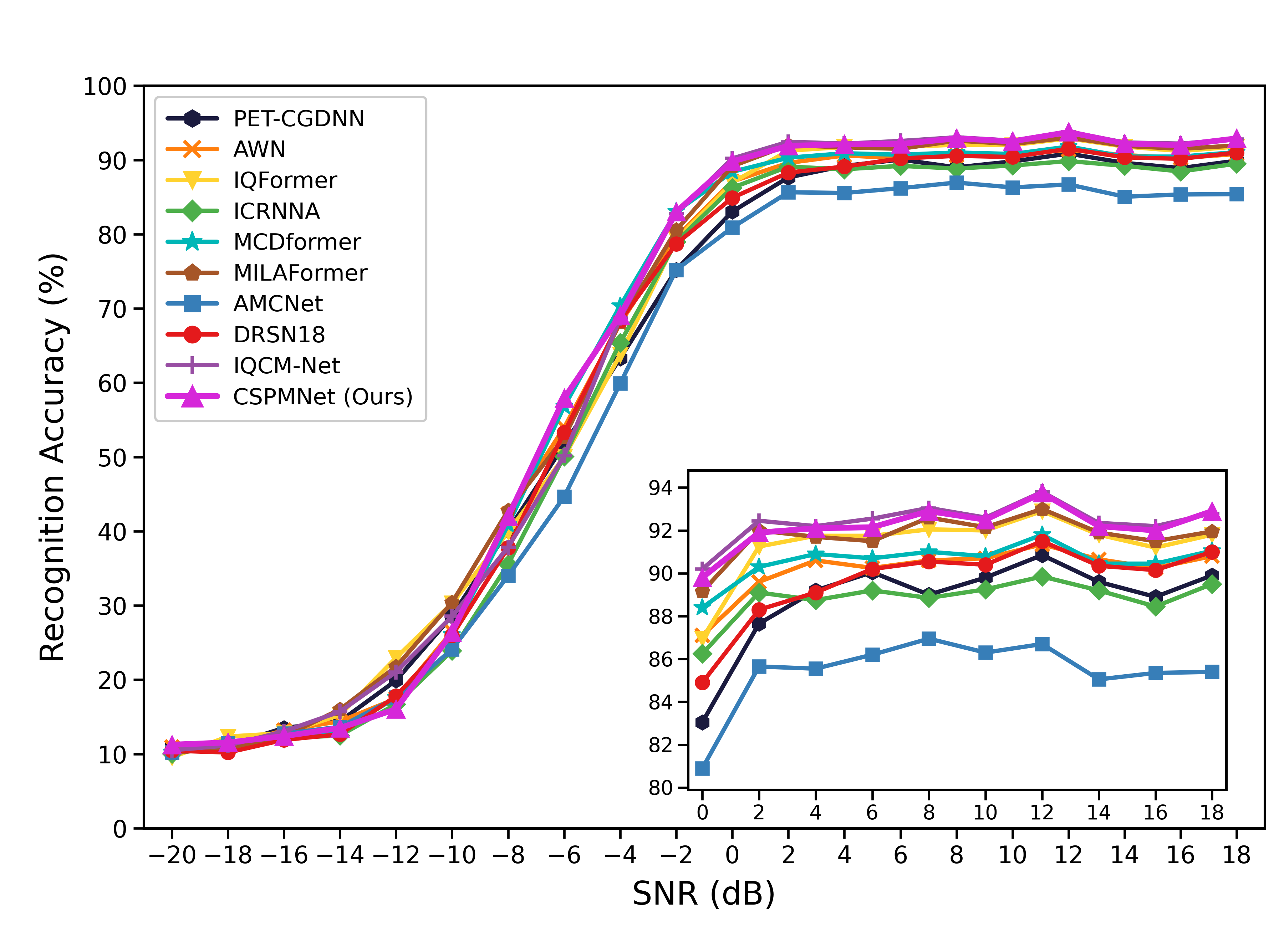}
        \centerline{(b) RadioML2016.10B}
    \end{minipage}
    \caption{Recognition accuracy of CSPMNet and other AMC models under different SNRs. (a) RadioML2022.01A. (b) RadioML2016.10B.}
    \label{fig:snr-curve}
\end{figure*}

To further evaluate robustness under different channel qualities, Fig.~\ref{fig:snr-curve} compares the recognition accuracy of CSPMNet and other AMC models at each SNR point.
As shown in Fig.~\ref{fig:snr-curve}(a), CSPMNet achieves the best accuracy at 20 of the 21 SNR points on RadioML2022.01A. 
The only exception is $-20$ dB, where it is slightly lower than DRSN18 by 0.45 percentage points. 
Compared with IQCM-Net, CSPMNet maintains positive gains at all SNR points, with the largest gain of 2.48 percentage points appearing at $-18$ dB. 
Compared with the lightweight AWN baseline, the largest gain appears at $-8$ dB and reaches 4.50 percentage points.

The advantage is most evident from the low-to-transition region to 0 dB. 
In this interval, noise is still strong enough to disturb modulation cues, but the signal is no longer completely unrecoverable.
This trend is consistent with the motivation of amplitude-preserving phase-motion modeling, where temporal phase dynamics become informative while magnitude reliability helps suppress weak noise-dominated responses.
As a result, CSPMNet achieves the best low-, mid-, and high-SNR averages on RadioML2022.01A, as reported in Table~\ref{tab:performance-complexity}.

Fig.~\ref{fig:snr-curve}(b) shows a more bounded trend on RadioML2016.10B.
CSPMNet does not dominate most individual SNR points, but it tracks the strongest curves from the transition region upward and achieves the best average accuracy over $-8$--0 dB.
In this mid-SNR interval, it exceeds IQCM-Net by 2.41 percentage points and MCDformer by 0.46 percentage points.
At high SNRs, CSPMNet remains close to IQCM-Net, while its low-SNR average is not the best.
Thus, the per-SNR curves support the same interpretation as Table~\ref{tab:performance-complexity}: CSPMNet is strongest on RadioML2022.01A and remains competitive on RadioML2016.10B without implying uniform pointwise dominance.

\subsubsection{Ablation Study}
To evaluate the contribution of the model design, we compare CSPMNet with three variants on both datasets.
As shown in Table~\ref{tab:frontend-ablation}, all variants are less accurate than CSPMNet on both datasets.
Using phase-motion features without learnable subband filtering, or replacing the free complex subband filters with fixed or constrained Morlet filters, consistently reduces OA.
This indicates that the proposed modules play a positive role in CSPMNet.

\begin{table}[t]
\caption{Ablation Study on RadioML2022.01A and RadioML2016.10B}
\label{tab:frontend-ablation}
\centering
\small
\setlength{\tabcolsep}{2.4pt}
\renewcommand{\arraystretch}{1.08}
\begin{tabular*}{\columnwidth}{@{\extracolsep{\fill}}lcc@{}}
\hline
\textbf{Variant} & \textbf{RadioML2022.01A} & \textbf{RadioML2016.10B} \\
\hline
PhaseMotion only & 61.36 ($\downarrow$ 9.73) & 59.30 ($\downarrow$ 3.96) \\
Fixed Morlet subband & 57.51 ($\downarrow$ 13.58) & 56.45 ($\downarrow$ 6.81) \\
Learnable Morlet subband & 69.01 ($\downarrow$ 2.08) & 57.68 ($\downarrow$ 5.58) \\
\textbf{CSPMNet} & \textbf{71.09} & \textbf{63.26} \\
\hline
\end{tabular*}
\end{table}

\section{Conclusion}
In this letter, we propose CSPMNet, a Complex Subband Phase-Motion Network for efficient AMC from raw I/Q samples. 
The proposed model enhances the accuracy-complexity tradeoff by incorporating learnable complex subband filtering and amplitude-preserving phase-motion modeling into a lightweight temporal classifier.
Experimental results on benchmark datasets demonstrate that CSPMNet, compared with existing AMC models, not only achieves highly competitive recognition performance but also benefits from substantially reduced model complexity, making it suitable for lightweight spectrum-monitoring receivers. 
Moreover, the proposed front end improves the extraction of modulation-discriminative subband dynamics across different SNR conditions. 
These results indicate that physics-guided complex front ends have practical potential for efficient AMC in resource-constrained communication scenarios.
Future work will explore adaptive complex subband modeling within AMC to further extend physics-guided representation learning toward more dynamic and heterogeneous spectrum environments.
\bibliographystyle{IEEEtran}
\bibliography{IEEEabrv,references}
\newpage

\end{document}